\documentclass[fleqn,usenatbib]{mnras}
\usepackage[latin1]{inputenc}
\usepackage[T1]{fontenc}
\usepackage{ae,aecompl}
\usepackage{graphicx}
\title[Constraints from growth-rate data on some dark...]{Constraints from growth-rate data on some coupled dark energy models mimicking a $\Lambda CDM$ expansion}
\author[Stéphane Fay]{Stéphane Fay\thanks{steph.fay@gmail.com}\\
Palais de la Découverte\\
Astronomy Department\\
Avenue Franklin Roosevelt\\
75008 Paris\\
France}
\date{Accepted XXX. Received YYY; in original form ZZZ}
\pubyear{2015}
\begin{document}
\label{firstpage}
\pagerange{\pageref{firstpage}--\pageref{lastpage}}
\maketitle
\begin{abstract}
The $\Lambda CDM$ expansion could be mimicked by a dark energy coupled to matter. Then, the equation of state $\bar w$ and coupling $\bar Q$ of this coupled dark energy could not be constrained by observations of the Hubble function alone. Also, in this paper, we determine the constraints on two such coupled dark energy models considering some current and forecast Euclid-like growth-rate data and assuming the prior on the $\Lambda CDM$ dark matter density parameter today $\Omega_{m0}=0.295\pm 0.04$. The first model is defined by a constant equation of state. We find that at $2\sigma$, $\bar w=-1.02_{-0.22}^{+0.06}$ and the coupling function $\bar Q_0$ today is $\bar Q_0H_0^{-3}=0.057_{-0.148}^{+0.353}$ with $H_0$ the Hubble constant. The second model is defined by a varying equation of state $\bar w=\bar w_a-\bar w_b\ln(1+z)$, with $z$ the redshift and $(\bar w_a,\bar w_b)$, two constants. We find that at $2\sigma$, $\bar w_a=-0.99_{-0.90}^{+0.17}$, $\bar w_b=-0.04_{-1.17}^{+0.31}$ and $\bar Q_0H_0^{-3}=0.0002_{-0.18}^{+1.35}$. These constraints on coupled dark energy agreed with a $\Lambda CDM$ model but are too poor to discard confidently a coupled dark energy different from vacuum but mimicking a $\Lambda CDM$ expansion.
\end{abstract}
\section{Introduction} \label{s0}
Degeneracy in cosmology takes several forms\citep{AviCer11,How12,Cro03,Wei08}. One of them is related with the cosmological models, different from the $\Lambda CDM$ one, but that mimics its expansion\citep{Fay07, Set13}. These models thus have the same Hubble function as the $\Lambda CDM$ model. This is the case for some dark energy coupled to dark matter models that we consider in this paper.\\
The possibility of a coupling between dark components has a long history that dates back before dark energy concept. Hence it was used to describe dark matter with varying mass in \citet{Gar93} or to solve the cosmological constant problem in \citet{Wet95}. Following the discovery of the cosmic acceleration\citep{Per99, Rie98}, interacting dark energy and its cosmological consequences were studied in \citet{Ame00A} or used to alleviate the coincidence problem in \citet{Toc02}. Many papers also looked for observational constraints on the coupling between dark energy and matter, e.g. \citet{Ame00,Oli05,Yan14,Cos14}.\\
In the present paper, we consider the possibility that such a coupling mimics a $\Lambda CDM$ expansion. Then, this coupling would not be detected by observations based only on the Hubble function such as distance-luminosity\citep{Per99, Rie98}, BAO peak\citep{Del14}, redshift drift\citep{Gen15}, etc. These observations can determine accurately the matter density parameter of the $\Lambda CDM$ model. However, they cannot determine if the coupled dark energy equation of state $\bar w$ is different from the one of the $\Lambda CDM$ model with a non vanishing coupling function $\bar Q$ between dark energy and dark matter. In a more physical viewpoint, the question is thus to know if the Universe could be described by a dark energy that is not a vacuum energy but mimics the $\Lambda CDM$ expansion thanks to its exchange with dark matter. One way to answer is to take into account observations based on growth-rate data(see for instance \citet{Toj12, Bla12}). Then, a coupled dark energy model mimicking the expansion of a $\Lambda CDM$ model cannot, generally, also mimic its growth-rate\citep{Hut15}. This last type of observations is thus able to distinguish between these two kinds of models.\\
In this paper, we use current growth-rate data and some forecast Euclid-like\citep{Lau11} data collected in \citet{Tad14,Ame14} to constrain two coupled dark energy models defined by their equations of state $\bar w$ and mimicking the $\Lambda CDM$ expansion. The form of $\bar w$ and this mimicry then set the form of the coupling function $\bar Q$ between dark energy and dark matter. Despite current growth-rate data are not very accurate, we show that they are able to constrain (poorly but in some finite confidence contours) $\bar w$, $\bar Q$ and the coupled dark matter density parameter today, $\bar\Omega_{m0}$, if we assume a prior on the $\Lambda CDM$ dark matter density parameter. These constraints are improved if we also take into account some forecast Euclid-like data\citep{Ame14}, in particular when the equation of state is varying. They are in agreement with a $\Lambda CDM$ model at $1\sigma$ although they still seem too poor to discard confidently a varying dark energy mimicking a $\Lambda CDM$ expansion.\\
The plan of the paper is the following. In the second section, we present a method to construct a coupled dark energy model having exactly the same expansion as the $\Lambda CDM$ model. In the third section, we present the differential equation for the dark matter density contrast when it is coupled with a dark energy\citep{Dev15} and the two above mentioned sets of growth-rate data\citep{Tad14,Ame14}. We check how they constrain the $\Lambda CDM$ model. In a fourth section, we determine the constraints on two coupled dark energy models mimicking the expansion of a $\Lambda CDM$ model but having a constant and a varying equation of state for dark energy. We conclude in the last section.
\section{Field equations and degeneracy between $\Lambda CDM$ and dark energy coupled models} \label{s1}
In this section, we show how a coupled dark energy model can mimic a $\Lambda CDM$ expansion. The quantities related to the coupled dark energy are indicated with a bar. We choose as unit $\frac{8\pi G}{3}=1$.\\
The equations for the coupled model are
\begin{equation}\label{H2}
\bar H^2=\bar\rho_m+\bar\rho_d
\end{equation}
\begin{equation}\label{rhom}
\bar\rho_m'+3\bar\rho_m=\bar Q/\bar H
\end{equation}
\begin{equation}\label{rhod}
\bar\rho_d'+3(1+\bar w)\bar\rho_d=-\bar Q/\bar H
\end{equation}
A prime means a derivative with respect to $N=\ln a$, with $a$ the scale factor of the $FLRW$ metric. $\bar H$ is the Hubble function. $\bar\rho_m$ and $\bar\rho_d$ are respectively the densities of dark matter and dark energy with an equation of state $\bar w$. The coupling between these densities is described by the coupling function $\bar Q$.\\
The equations for the non coupled model are
\begin{equation}\label{H21}
H^2=\rho_m+\rho_d
\end{equation}
\begin{equation}\label{rhom1}
\rho_m'+3\rho_m=0\Rightarrow \rho_m=\rho_{m0}e^{-3N}
\end{equation}
\begin{equation}\label{rhod1}
\rho_d'+3(1+w)\rho_d=0
\end{equation}
The quantities without the bar have the same meaning as the quantities with the bar for the coupled model. $\rho_{m0}$ is the non coupled matter density today. The $6$ above equations contain $3$ unknowns that we fix by choosing $\bar w$, $w$ (that is $w=-1$ when considering the $\Lambda CDM$ model) and $\bar H=H$ such that the coupled dark energy model mimics the expansion of the non coupled model. We want to determine $\bar Q$ as a function of $\bar w$, $w$ and $\bar H$. To reach this goal, we define the difference $\Delta$ between the coupled and non coupled matter densities
\begin{equation}\label{rhoRed}
\Delta=\bar\rho_m-\rho_m
\end{equation}
Replacing $\bar\rho_m$ in (\ref{rhom}) with $\Delta$ and taking into account (\ref{rhom1}), it comes
\begin{equation}\label{rhom2}
\Delta'+3\Delta=\bar Q/\bar H
\end{equation}
Moreover, the Hubble function $\bar H$ rewrites as
$$
\bar H^2=\rho_m+\Delta+\bar\rho_{d}
$$
Comparing $\bar H$ to the Hubble function $H$, it comes since $\bar H= H$
\begin{equation}\label{rhod1formal}
\rho_d=\Delta+\bar\rho_{d}
\end{equation}
Then, summing (\ref{rhod}) with (\ref{rhom2}) and using (\ref{rhod1formal}), we get
\begin{equation}\label{rhod1w1}
\rho_d'+3\frac{\left[(1+\bar w)\bar\rho_d+\Delta\right]}{\Delta+\bar\rho_{d}}\rho_d=0
\end{equation}
Comparing this last relation with (\ref{rhod1}), we thus obtain 
\begin{equation}\label{w1}
1+ w=\frac{\left[(1+\bar w)\bar\rho_d+\Delta\right]}{\Delta+\bar\rho_{d}}
\end{equation}
that rewrites as
\begin{equation}\label{rhodw1}
\bar\rho_d=\frac{w \Delta}{\bar w-w}
\end{equation}
Then, subtracting the two Hubble functions (\ref{H2}) and (\ref{H21}) and using (\ref{rhoRed}) and (\ref{rhodw1}), we get
\begin{equation}\label{rhom2ww1}
\Delta=\frac{(\bar w-w)\rho_d}{w}
\end{equation}
This last expression allows to replace $\Delta$ in (\ref{rhom2}) to finally get
\begin{equation}\label{Q}
\bar Q=-\frac{\bar H \rho_d \left[3 \bar w^2 w- w\bar w'+\bar w\left(-3 w^2+w'\right)\right]}{\bar w^2}
\end{equation}
Hence, when we choose $\bar w$ and $w$, we can calculate $\rho_d$ from (\ref{rhod1}) and then $\bar H= H$ from (\ref{H21}), thus defining completely the non coupled model. Then, we can get $\bar Q$ from (\ref{Q}), $\Delta$ from (\ref{rhom2ww1}), $\bar \rho_m$ from (\ref{rhoRed}) and $\bar \rho_d$ from (\ref{rhodw1}), thus defining completely the coupled model.\\
Concerning the integration constants $\bar \rho_d(0)$ and $\bar \rho_m(0)$, they are determined by rewriting equation (\ref{w1}) with help of equation (\ref{rhod1formal}) as
\begin{equation}\label{relD}
1+w=\frac{\bar w \bar\rho_d+\rho_d}{\rho_d}
\end{equation}
Observations like supernovae impose the value of $\rho_d(0)=H_{0}^2(1-\Omega_{m0})$ with $H_{0}$ the Hubble constant that in this paper is $H_0=\bar H_0=70km/s/Mpc$ and $\Omega_{m0}$ the density parameter of the non coupled dark matter today. Hence, when we choose $\bar w(N)$ and $w(N)$, the above equation defines $\bar\rho_d(0)$ in $N=0$, the value of the coupled dark energy today and thus the integration constant in (\ref{rhod}). Then we get the value of the coupled dark matter today $\bar\rho_m(0)$ thanks to the Hubble function (\ref{H2}).\\\\
In the rest of this section, we consider the special case of a coupled dark energy mimicking a $\Lambda CDM$ expansion, i.e. $w=-1$. We thus obtain from (\ref{relD}) in $N=0$
$$
\frac{\rho_d(0)}{\bar\rho_d(0)}=-\bar w(0)
$$
Then, considering the density parameters for matter ($\Omega_{m0}$, $\bar \Omega_{m0}$) and dark energy ($\Omega_{d0}$, $\bar \Omega_{d0}$) and taking into account the constraints $\Omega_{d0}=1-\Omega_{m0}$ and $\bar\Omega_{d0}=1-\bar\Omega_{m0}$, it comes
\begin{equation}\label{rel0}
\bar\Omega_{m0}=\frac{1+\bar w(0)-\Omega_{m0}}{\bar w(0)}
\end{equation}
This relation allows to determine some constraints on $\bar\Omega_{m0}$ when assuming some values for $\Omega_{m0}$ coming from supernovae observations and when deriving some constraints on $\bar w(0)$ coming from growth rate data (that will be done in section \ref{s3}). Since $\Omega_{m0}<1$, it shows that if $\bar w(0)<-1$, then $\bar\Omega_{m0}>\Omega_{m0}$: there is more matter in the coupled model than in the non coupled one at present time if the coupled dark energy is presently a ghost. The opposite is true when $\Omega_{m0}<1$ and $-1<\bar w(0)<-1/3$, i.e. when the coupled dark energy is quintessence. These remarks that apply to present time can be extended to any $N$ by considering equation (\ref{w1}) that rewrites when $w=-1$
$$
(1+\bar w)\bar\rho_d=-\Delta
$$
Assuming that the coupled dark energy density $\bar\rho_d$ is positive, it follows that if $\bar w<-1$, then $\bar\rho_m>\rho_m$ and $\bar\rho_m<\rho_m$ otherwise. Physically, this means that if the expansion of the $\Lambda CDM$ model is mimicked by a coupled dark energy, when the dark matter density of the coupled model is larger (smaller) than the one predicted by the standard $\Lambda CDM$ model, the coupled dark energy is a ghost (respectively quintessence). Hence, the crossing of the phantom divide $\bar w=-1$ corresponds to a coupled dark matter density becoming larger or smaller than the dark matter density of the $\Lambda CDM$ model. Such a crossing can be in agreement with the data as shown in subsection \ref{s22}.\\
Such a link between the sign of $\bar w+1$ and the quantity of coupled dark matter is also recovered in the expression (\ref{Q}) for $\bar Q$ that rewrites with $w=-1$
$$
\bar Q=\frac{\bar H \rho_d (3\bar w+3\bar w^2-\bar w')}{\bar w^2}
$$
$\rho_d$ is the constant vacuum energy. Then, when $\bar w'<<(\bar w,\bar w^2)$, i.e. the coupled dark energy equation of state does not vary too much, and still assuming that $\rho_d>0$, the sign of $\bar Q$ is the one of $(\bar w+1)/\bar w$. Hence, when the coupled dark energy is a ghost ($\bar w<-1$), dark energy is cast into matter since $\bar Q>0$ whereas when the coupled dark energy is quintessence ($-1<\bar w<-1/3$), matter is cast into dark energy since $\bar Q<0$. Some examples of coupling functions for some specific forms of $\bar w$ are plotted as functions of the redshift in section \ref{s3}.
\section{Dark matter density contrast and data} \label{s2}
\begin{figure}
\centering
\includegraphics[width=6cm]{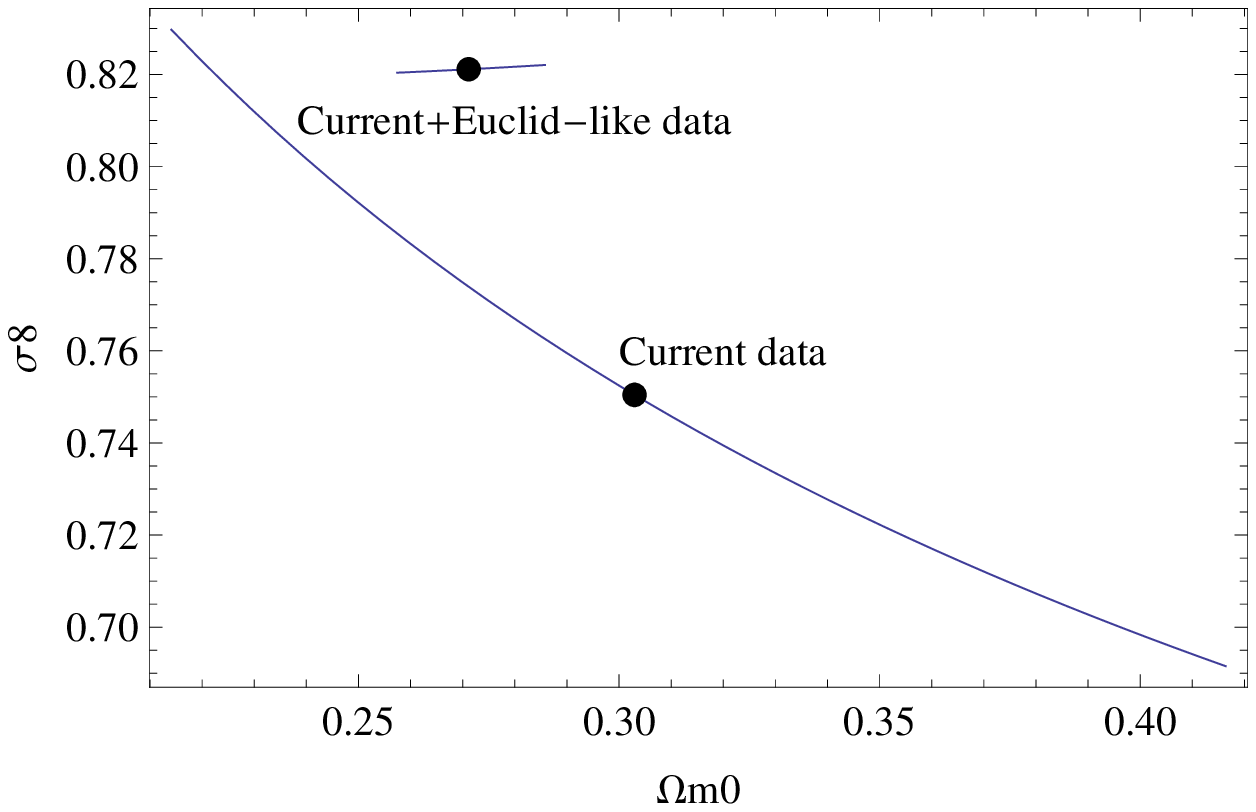}
\includegraphics[width=6cm]{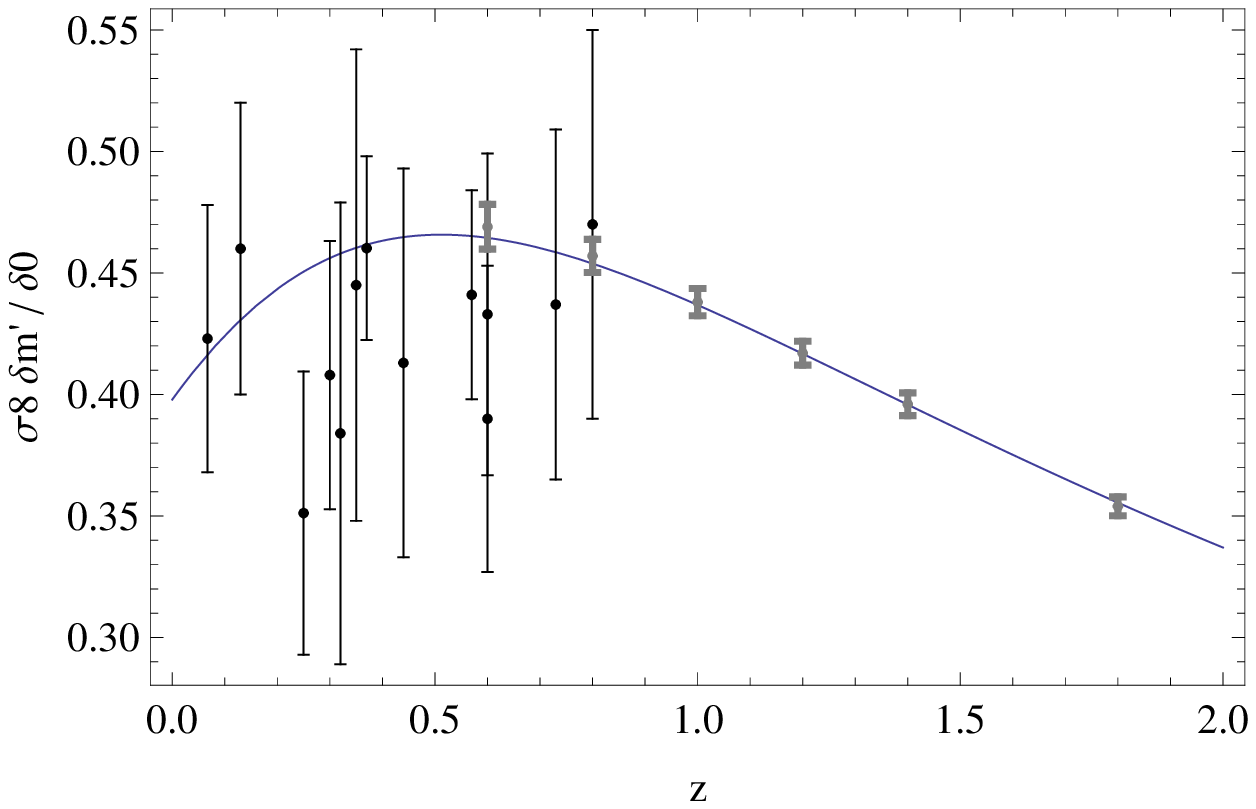}
\caption{\scriptsize{\label{fig1}First graph: $(\Omega_{m0},\sigma_8)$ at $1\sigma$ for the $\Lambda CDM$ model when considering current growth-rate data alone and with some forecast Euclid-like data. The points indicate the best fits. Second graph: Best fit for $d$ with the current growth-rate data (black) and some forecast Euclid-like data (thick gray) for the $\Lambda CDM$ model.}}
\end{figure}

When dark energy is coupled to dark matter, the evolution equation for the dark matter density contrast $\delta_m=\delta\bar \rho_m/\bar \rho_m$, with $\delta\bar \rho_m$ the dark matter perturbations, writes\citep{Ame04,Bor08,Dev15}
\begin{equation}\label{pert}
\ddot\delta_m+\dot\delta_m(2\bar H+\frac{\bar Q}{\bar\rho_m})+\delta_m(-\frac{3}{2}\bar\rho_m+2\frac{\bar H \bar Q}{\bar\rho_m}-\frac{\bar Q\dot{\bar\rho_m}}{\bar\rho_m^2}+\frac{\dot{\bar Q}}{\bar \rho_m})=0
\end{equation}
where a dot means a derivative with respect to $t$. The $\Lambda CDM$ model corresponds to $\bar Q=0$. Following \citet{Tad14}, we consider some initial conditions in $N_i=-1.5$, i.e. at the redshift $z_i=e^{-N_i}-1=3.48$
$$
\delta_{m_i}=e^{N_i}
$$
and
$$
d\delta_{m_i}/dN_i=\alpha e^{N_i}
$$
$\alpha$ is a constant. For sake of simplicity we consider the special\footnote{In \citet{Tad14}, $\alpha$ is either chosen as $\alpha=1$ or as a free parameter.} value $\alpha=1$. We checked that our results are insensitive to the value of $N_i$. This is due to the fact that, for the models we study in this paper, $\delta_{m}\simeq e^{N}$ is a good approximation for redshift larger than $2$ and when matter is dominating\citep{Con13}. The growth rate $d$ is defined as
$$
d=\sigma_8\frac{\delta'}{\delta_0}
$$
where $\sigma_8$ is the present $(N=0)$ power spectrum normalisation and $\delta_0$ the present dark matter density contrast. In the following, we will use current growth-rate data and some forecast Euclid-like data issued from \citet{Tad14,Ame14}. They are presented in table \ref{tab1} and plotted on the second graph of figure \ref{fig1}.
\begin{table}
\centering
\caption{Current growth-rate data (left) and some forecast Euclid-like data (right) $d$ at redshift $z$ with error $\sigma$ issued from \citet{Tad14,Ame14}.}
\label{tab1}
\begin{tabular}{ccc|c|ccc}
\hline
 &Current& & & &Forecast&\\
$z$ & $d$ & $\sigma$ & &$z$ & $d$ & $\sigma$\\
\hline
 0.067 & 0.423 & 0.055 &&0.6  & 0.469  & 0.0092 \\
 0.25 & 0.3512 & 0.0583 &&0.8  & 0.457  & 0.0068\\
 0.37 & 0.4602 & 0.0378 &&1.   & 0.438  & 0.0056\\
 0.3  & 0.408  & 0.0552 &&1.2  & 0.417  & 0.0049\\
 0.6  & 0.433  & 0.0662 &&1.4  & 0.396  & 0.0047\\
 0.44 & 0.413  & 0.08 &&1.8  & 0.354  & 0.0039\\
 0.6  & 0.39   & 0.063 &&&&\\
 0.73 & 0.437  & 0.072 &&&&\\
 0.8  & 0.47   & 0.08 &&&&\\
 0.13 & 0.46   & 0.06 &&&&\\
 0.35 & 0.445  & 0.097 &&&&\\
 0.32 & 0.384  & 0.095 &&&&\\
 0.57 & 0.441  & 0.043 &&&&\\
 \hline
\end{tabular}
\end{table}
To constrain a cosmological model, we minimize the following $\chi^2$
$$
\chi^2=\sum _{k=1}^n \frac{(d(z_k)-\sigma_8\frac{\delta'(z_k)}{\delta_0})^2}{\sigma(z_k)^2}
$$
where $d(z_k)$ are the observational data at redshift $z_k$ and $\sigma(z_k)$ their errors. We need to marginalise $\sigma_8$. This is done by looking for the value $\sigma_{8min}$ of $\sigma_8$ minimising $\chi^2$, i.e. $d\chi^2/d\sigma_8(\sigma_{8min})=0$. We find
$$
\sigma_{8min}=\frac{\sum _{j=1}^n d(z_j) \frac{\delta'(z_j)}{\delta_0 \sigma(z_j)^2}}{\sum _{j=1}^n \frac{\delta'(z_j)^2}{\delta_0^2\sigma(z_j)^2}}
$$
We then replace $\sigma_8$ by $\sigma_{8min}$ in $\chi^2$. We check this new definition of $\chi^2$ with the $\Lambda CDM$ model. Then, we find at $1\sigma$ with current growth-rate data that the best fit is got with a $\Lambda CDM$ dark matter density parameter $\Omega_{m0}=0.30^{-0.09}_{+0.12}$ with $\sigma_8=0.75^{-0.06}_{+0.08}$. If we also consider the forecast Euclid-like data, we get this time $\Omega_{m0}=0.27^{-0.01}_{+0.01}$ with $\sigma_8=0.82^{-0.01}_{+0.01}$. $1\sigma$ results in the $(\Omega_{m0},\sigma_8)$ space are shown on the first graph of figure \ref{fig1} and the best fit for $d$ is shown on the second graph. Euclid-like data improve the determination of the $\Omega_{m0}$ and thus $\sigma_8$ parameters. For comparison, Planck results from Sunyaev-Zeldovitch cluster counts give $\Omega_{m0}=0.29\pm 0.02$ with $\sigma_8=0.77\pm 0.01$ \citep{Ade14}.\\
One remarks that the best fitting value for $\Omega_{m0}$ obtained with current growth-rate data is not exactly the same when we also consider the Euclid-like forecast data. This is also the case for the free parameters of the two models we consider in section \ref{s3}. This does not mean that there is an inconsistency between the best fitting values of $\Omega_{m0}$ got with or without Euclid-like data. Firstly, the best fitting value obtained with Euclid-like data is in the $1\sigma$ interval of the fitting values got without Euclid-like data. Secondly, the differences between the best fitting values of $\Omega_{m0}$ are due to the fact that current and Euclid-like data are very different. The current growth-rate data are inhomogeneous (there is a large dispersion of these data as shown on the second graph of figure \ref{fig1}), they come from several surveys (BOSS, WiggleZ, etc, see \citet{Tad14} for a complete list) and they have large error bars. The forecast Euclid-like data are homogeneous (they are evaluated with a fiducial flat $\Lambda CDM$ model\citep{Ame14} characterised by the WMAP 7-year values) and have small error bars. Adding to the current data more data points with smaller error bars and less dispersion like the ones of Euclid-like data thus improves the cosmological parameters determination in two ways: it sets more accurately the value of $\Omega_{m0}$ than with the current observations alone (or other parameters for other cosmological models) and it shrinks the confidence contours got with these last data. The same remarks applied to the parameters of the models of subsections \ref{s21} and \ref{s22} that we determine similarly.\\
Finally, a last remark is related to an internal degeneracy of dark energy coupled models mimicking a $\Lambda CDM$ expansion (i.e. $w=-1$) when their equation of state $\bar w$ is such that $\bar w<<-1$ and $\bar w'/\bar w^2\simeq 0$. Then $\bar Q\simeq 3\bar H \rho_d$ and when we introduce this form of $\bar Q$ in equation (\ref{pert}), we can calculate the best $\chi^2$ of such a theory. It then depends on two parameters $\bar \Omega_{m0}$ and $\bar w(0)$ (that is introduced when using equation (\ref{rel0}) to replace $\Omega_{m0}$). With current growth rate data the best $\chi^2$ is found when $\bar \Omega_{m0}=0.95$ and $\bar w(0)=-20$. The $2\sigma$ confidence contour in the $(\bar \Omega_{m0},\bar w(0))$ then looks like a line along $\bar\Omega_{m0}\simeq 1$ when $\bar w(0)\rightarrow -\infty$. This degeneracy, that is also present when considering the Euclid-like data, thus allows to $\bar w(0)$ to diverge negatively when $\bar \Omega_{m0}\simeq 1$ although the model is still in agreement with the data. In section \ref{s3}, we show how to remove it by considering some observational constraints on $\Omega_{m0}$.
\section{Constraints on two coupled dark energy models} \label{s3}
In this section, we constrain two coupled dark energy models mimicking the $\Lambda CDM$ expansion ($w=-1$) with the growth-rate data presented on table \ref{tab1}. The first one is defined by a constant equation of state $\bar w$ and the other one by a linear equation of state $\bar w=\bar w_a+\bar w_b N$.
\subsection{$w=-1$ and $\bar w=const$} \label{s21}
\begin{figure}
\centering
\includegraphics[width=6cm]{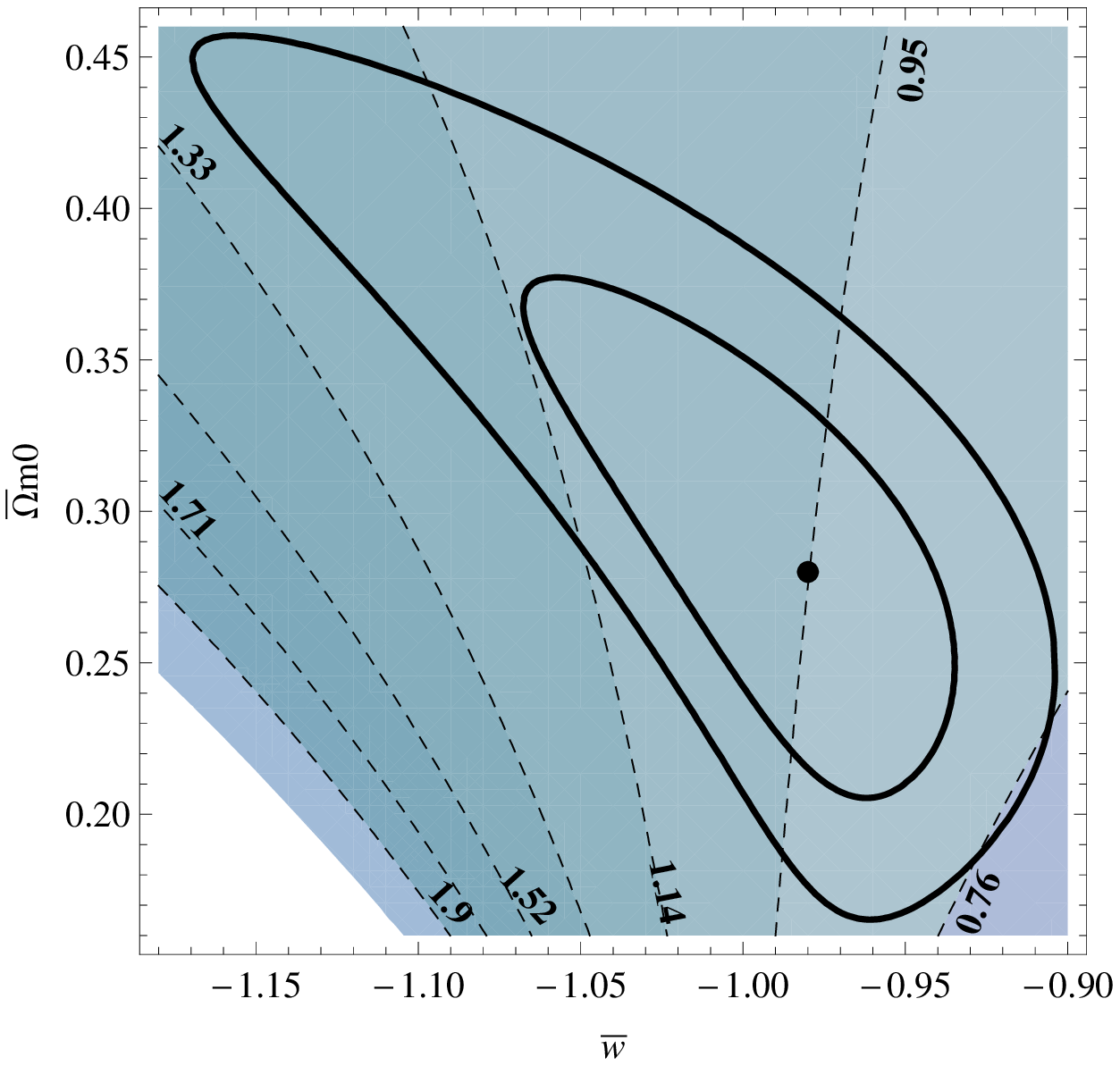}
\includegraphics[width=6cm]{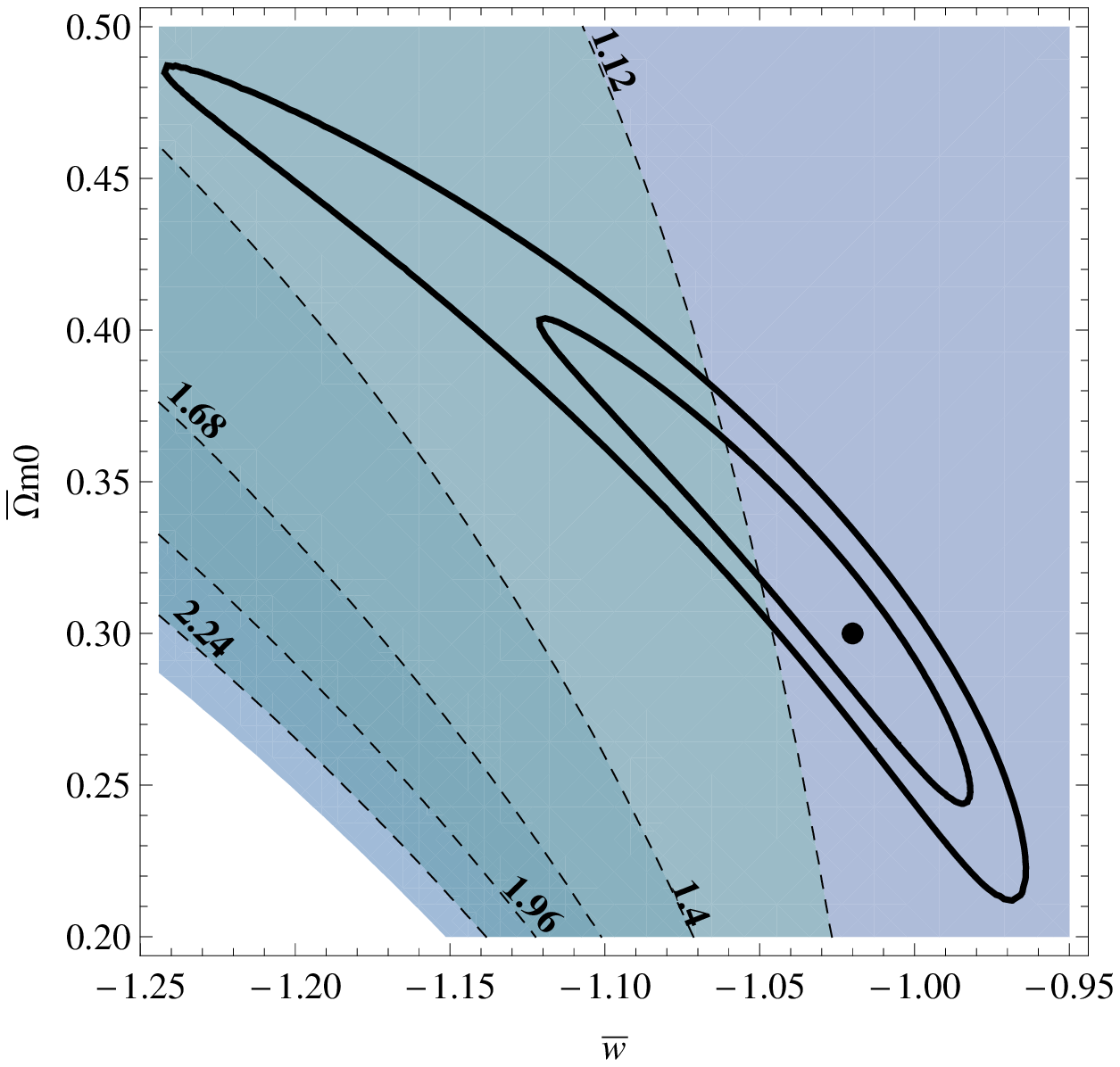}
\caption{\scriptsize{\label{fig2}$1$ and $2\sigma$ confidence contours obtained with current growth-rate data (first graph) and considering also the the forecast Euclid-like data (second graph). Lines with numbers give the ratio $\bar\Omega_{m0}/\Omega_{m0}$. Black dots give the best fit.}}
\end{figure}

\begin{figure}
\centering
\includegraphics[width=7cm]{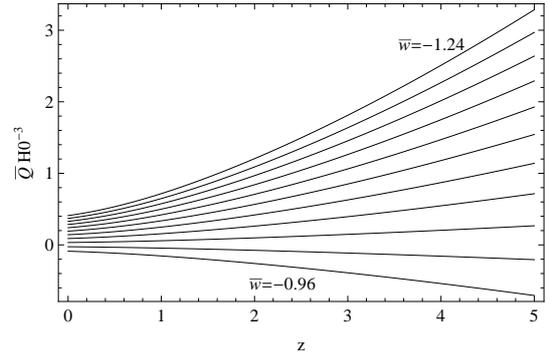}
\caption{\scriptsize{\label{fig6}Some coupling functions $\bar Q/H_0^3$ for some values of $\bar w$ as a function of the redshift $z$.}}
\end{figure}

We consider a coupled dark energy with a constant equation of state $\bar w$. Following the results of section \ref{s1}, this model mimics the expansion of a $\Lambda CDM$ model with $w=-1$ when 
$$
\bar\rho_m=\frac{H_0^2 (1+\bar w) (1-\Omega_{m0})}{\bar w}+e^{-3 N} H_0^2 \Omega_{m0}
$$
$$
\bar\rho_d= -\frac{H_0^2 (1-\Omega_{m0})}{\bar w}
$$
$$
\bar Q=\frac{3 H_0^2 (1 + \bar w) (1-\Omega_{m0})H}{\bar w}
$$
Moreover, from (\ref{rel0}) we derive that the coupled dark matter density parameter today $\Omega_{m0}$ writes
\begin{equation}\label{rel1}
\bar\Omega_{m0}=\frac{1+\bar w-\Omega_{m0}}{\bar w}
\end{equation}
Note that although the coupled dark energy $\bar\rho_d$ is a constant, its equation of state is not $-1$ since $\bar Q\not = 0$. As indicated in section \ref{s2}, the model $w=-1$ and $\bar w=const$ has an internal degeneracy that allows large values of $\bar w$ to be in agreement with the data when $\bar\Omega_{m0}\simeq 1$. It can be removed by taking into account the prior $\Omega_{m0}=0.295\pm 0.04$. This last value is observationally determined with supernovae data from Union 2.1 in \citet{Suz12} for the $\Lambda CDM$ model. It is thus independent from growth-rate data. This prior consists in adding to $\chi^2$ the term $(\Omega_{m0}-0.295)^2/0.04^2=(1+\bar w-\bar w\bar\Omega_{m0}-0.295)^2/0.04^2$. Then, if $\bar w$ is large and $\bar\Omega_{m0}\simeq  1$, this tends to increase $\chi^2$ and thus to discard large values of $\bar w$ from the two sigma interval.\\
Then, when considering current growth-rate data and the above prior on $\Omega_{m0}$, we get at $2\sigma$, $\bar\Omega_{m0}=0.28_{-0.12}^{+0.17}$ and $\bar w=-0.98_{-0.19}^{+0.08}$. These constraints are slightly improved if we also consider the forecast Euclid-like data. Then, we obtain at $2\sigma$, $\bar\Omega_{m0}=0.30_{-0.09}^{+0.18}$ and $\bar w=-1.02_{-0.22}^{+0.06}$. We also derive for the value of the coupling function today, $\bar Q(0)=\bar Q_0$, that $\bar Q_0 H_0^{-3}=0.057_{-0.148}^{+0.353}$. The confidence contours for $(\bar\Omega_{m0},\bar w)$ are plotted on figure \ref{fig2} with the ratio $\bar\Omega_{m0}/\Omega_{m0}$.\\
Finally, on figure \ref{fig6}, we plot some coupling functions $\bar Q/H_0^3$ for some values of $\bar w$ in agreement with these last constraints. As noted at the end of section \ref{s1}, since $\bar w'=0$, the sign of the coupling function from which depends the matter/dark energy transformation is the one of $(\bar w+1)/\bar w$: dark energy is cast into matter when it is a ghost and the opposite when it is quintessence. Moreover, as indicated by the form of $\bar Q$, the coupling function is an increasing function of the redshift when $\bar w<-1$ and a decreasing function when $-1<\bar w<0$. Hence, more and more dark matter (respectively dark energy) is cast into dark energy (respectively dark matter) when we go to the past and the coupled dark energy is a ghost (respectively quintessence).
\subsection{$w=-1$ and $\bar w=\bar w_a+\bar w_b N$} \label{s22}
\begin{figure}
\centering
\includegraphics[width=6cm]{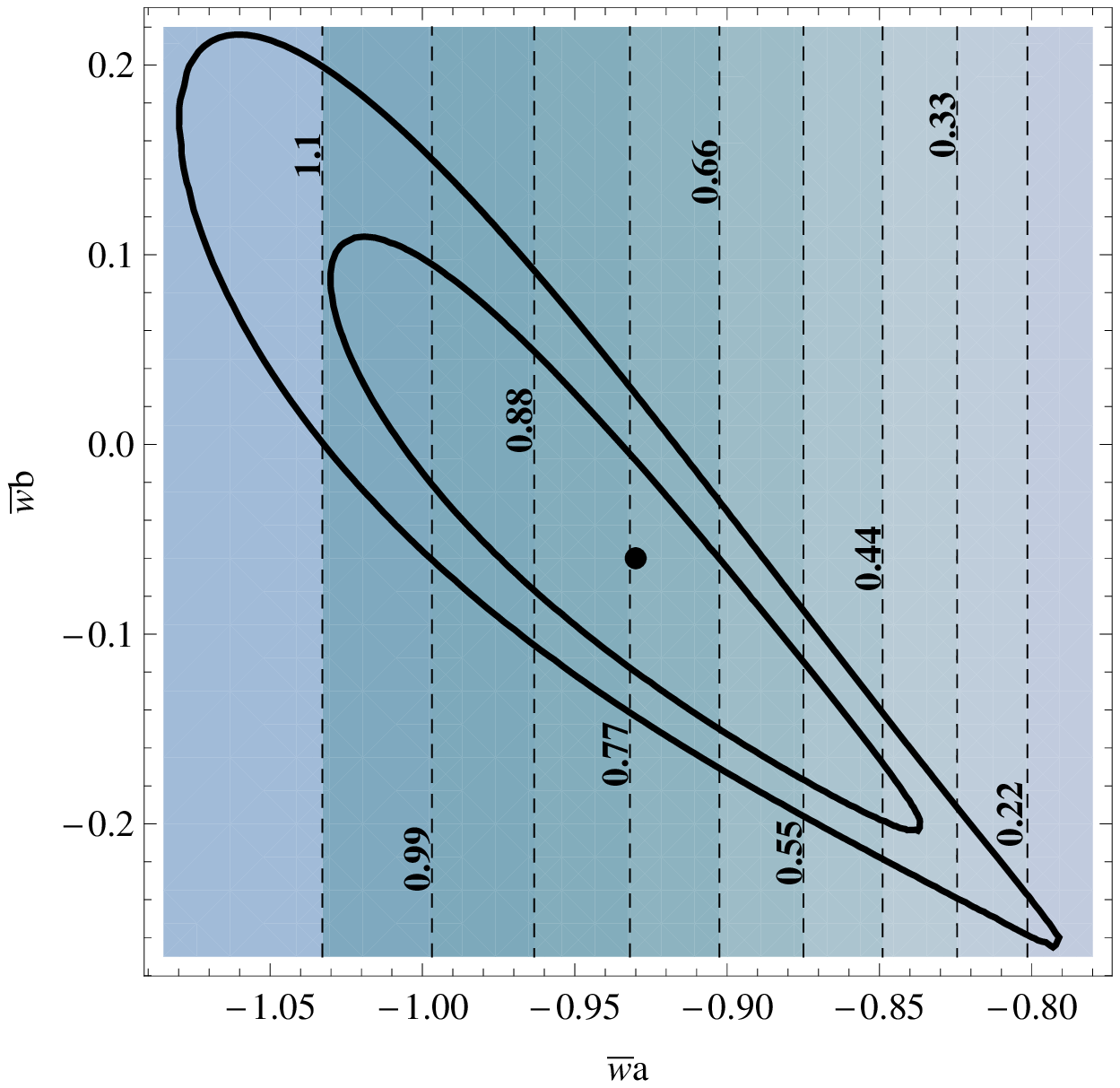}
\includegraphics[width=6cm]{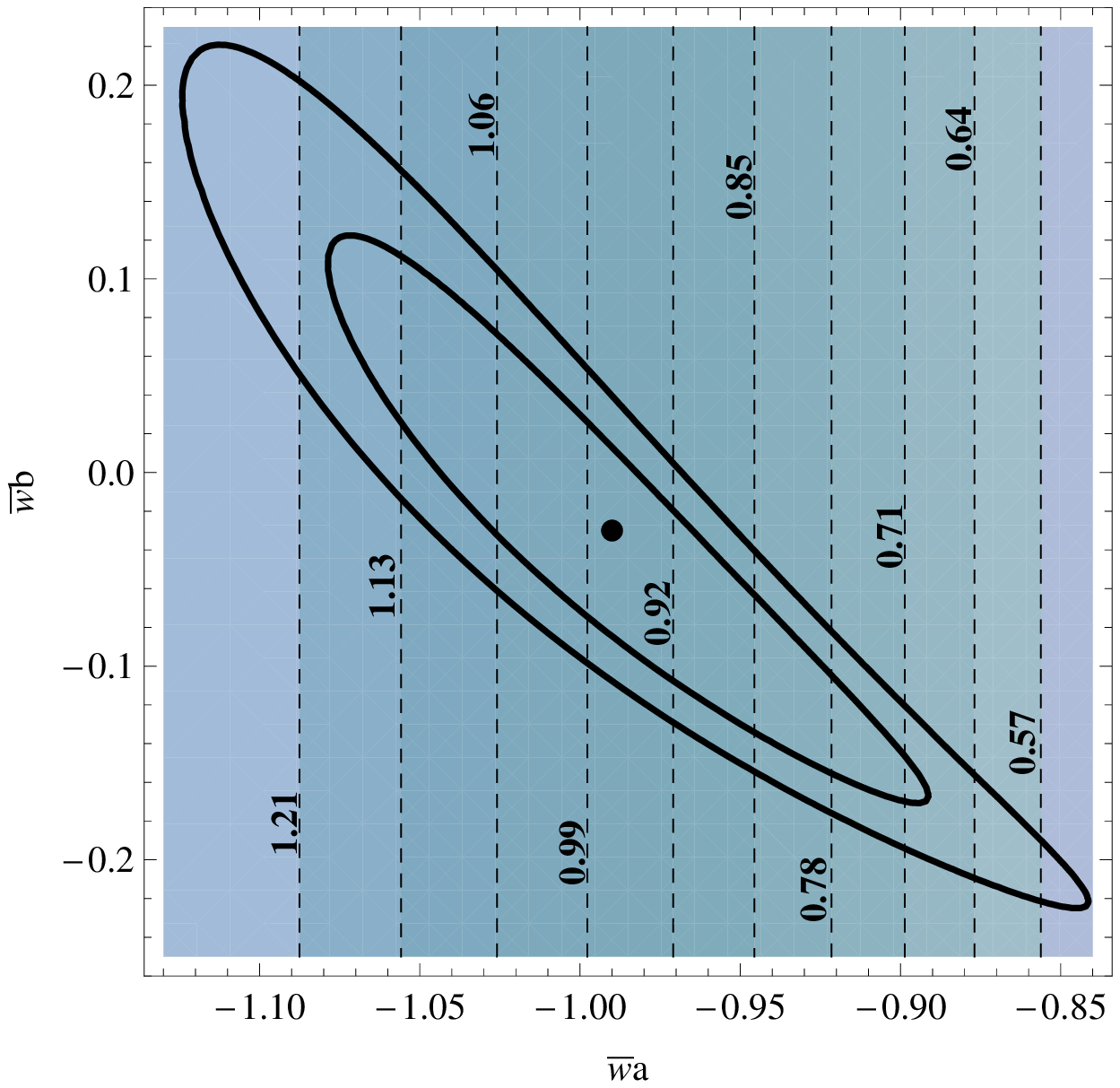}
\caption{\scriptsize{\label{fig3}$1$ and $2\sigma$ confidence contours for the best fitted values of $\bar\Omega_{m0}$ obtained with current growth-rate data (first graph when $\bar\Omega_{m0}=0.24$) and considering also the forecast Euclid-like data (second graph when $\bar\Omega_{m0}=0.28$). Lines with numbers correspond to the ratio $\bar\Omega_{m0}/\Omega_{m0}$. Black dots give the best fits.}}
\end{figure}

\begin{figure}
\centering
\includegraphics[width=7cm]{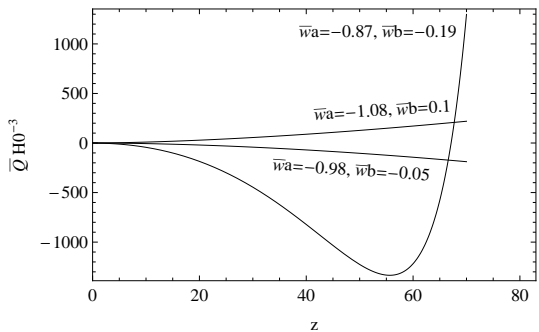}
\caption{\scriptsize{\label{fig7}Some coupling functions $\bar Q H_0^{-3}$ for some values of $\bar w$ in agreement with current growth-rate and forecast Euclid-like data.}}
\end{figure}

We consider a varying equation of state $\bar w=\bar w_a+\bar w_b N$. The coupled model mimicking the expansion of the $\Lambda CDM$ model is then defined by
$$
\bar\rho_m=\frac{H_0^2 (1+\bar w_a+\bar w_b N )(1-\Omega_{m0})}{\bar w_a+\bar w_b N}+e^{-3 N} H_0^2 \Omega_{m0}
$$
$$
\bar\rho_d=-\frac{H_0^2 (1-\Omega_{m0})}{\bar w_a+ \bar w_b N}
$$
$$
\bar Q=\frac{H_0^2 \left[-\bar w_b+3 (\bar w_a+\bar w_b N)+3 (\bar w_a+\bar w_b N)^2\right] (1-\Omega_{m0})H}{(\bar w_a+\bar w_b N)^2}
$$
Moreover, from (\ref{rel0}) we derive that the coupled dark matter density parameter today is
$$
\bar\Omega_{m0}=\frac{1+\bar w_a-\Omega_{m0}}{\bar w_a}
$$
For the same reasons as in subsection \ref{s21}, we still assume the prior $\Omega_{m0}=0.295\pm 0.04$. Then, considering only current growth-rate data, we get at $2\sigma$, $\bar w_a=-0.93_{-7.36}^{+0.14}$, $\bar w_b=-0.06_{-5.44}^{+3.88}$ and $\bar\Omega_{m0}=0.24_{-0.12}^{+0.67}$ that is quite bad. If we also consider the forecast Euclid-like data, we get at $2\sigma$, $\bar w_a=-0.99_{-0.90}^{+0.17}$ and $\bar w_b=-0.04_{-1.17}^{+0.31}$. Hence, Euclid-like data clearly improve the constraints on the equation of state. We then also derive that today $\bar\Omega_{m0}=0.28_{-0.09}^{+0.33}$ and $\bar Q_0 H_0^{-3}=0.0002_{-0.18}^{+1.35}$. Some confidence contours for $(\bar w_a,\bar w_b)$ for the best fitted values $\bar\Omega_{m0}=0.24$ and $0.28$ and with the ratio $\bar\Omega_{m0}/\Omega_{m0}$ are plotted on figure \ref{fig3}.\\
Finally, let us say some few words about the properties of $\bar w$ and $\bar Q$. Obviously, the equation of state $\bar w_a+\bar w_b N$ crosses the line $-1$ for a finite value of $N<0$ (i.e. in the past) when $\bar w_a+1$ and $\bar w_b$ have the same sign. Moreover, $\bar Q$ diverges for a finite value of $N<0$ when $\bar w_a$ and $\bar w_b$ have the same sign. None of these possibilities is excluded by the data. To avoid the crossing of the line $\bar w=-1$ and the divergence of $\bar Q$, we then need that $\bar w_a<-1$ and $\bar w_b>0$ or $\bar w_a>0$ and $\bar w_b<0$. Only the first possibilities agrees with the data. This is shown at $1\sigma$ on figure \ref{fig3} for some special values of $\bar\Omega_{m0}$. We plot some coupling functions $\bar Q H_0^{-3}$ on figure \ref{fig7} for some values of $\bar w_a$ and $\bar w_b$ in agreement with current growth-rate and forecast Euclid-like data when $\bar\Omega_{m0}=0.28$, including a diverging coupling function. 
\section{Conclusion} \label{s4}
The standard model of cosmology is the $\Lambda CDM$ model. In this paper we examined if a dark energy different from a vacuum energy but coupled to dark matter and mimicking a $\Lambda CDM$ expansion could also describe our Universe. To reach this goal, we first explained how to define a coupled dark energy model mimicking a $\Lambda CDM$ expansion. Then since observational data related to Universe expansion cannot discriminate between a $\Lambda CDM$ model and such a coupled dark energy model, we are led to use growth-rate data since a coupled dark energy mimicking a $\Lambda CDM$ expansion cannot generally\citep{Lom15} also mimic its growth-rate.\\
We then constrained two dark energy models, one with a constant equation of state $\bar w$ and the other one with a $\bar w$ varying linearly with respect to $N$. We use the prior $\Omega_{m0}=0.295\pm 0.04$ to remove an internal degeneracy that plagues coupled dark energy models mimicking a $\Lambda CDM$ expansion.\\
Then, we find at $2\sigma$ that a constant equation of state in agreement with current and forecast growth-rate data is such that $\bar w=-1.02_{-0.22}^{+0.06}$, the coupled dark matter density parameter is $\bar\Omega_{m0}=0.30_{-0.09}^{+0.18}$ and the value of the coupling function today is $\bar Q_0 H_0^{-3}=0.057_{-0.148}^{+0.353}$. If now we consider a varying equation of state $\bar w=\bar w_a+\bar w_bN$, we obtain that $\bar w_a=-0.99_{-0.90}^{+0.17}$, $\bar w_b=-0.04_{-1.17}^{+0.31}$, $\bar\Omega_{m0}=0.28^{+0.33}_{-0.09}$ and $\bar Q_0H_0^{-3}=0.0002_{-0.18}^{+1.35}$.\\
These two models that mimic a $\Lambda CDM$ expansion are thus, also from the viewpoint of growth rate data, in agreement with a $\Lambda CDM$ model, even at $1\sigma$. However the data are not (and should not be with Euclid) accurate enough to discard confidently the possibility of a Universe described by a coupled dark energy with a varying equation of state, despite a strong prior on $\Omega_{m0}$. Better and higher redshift data will be necessary to improve the constraints\citep{Seo14} on this special class of dark energy models able to mimic the $\Lambda CDM$ expansion.
\section*{Acknowledgment}
I thank the anonymous referee for his/her helpful comments.
\bibliographystyle{unsrt}

\begin{thebibliography}{10}
\bibitem[\protect\citeauthoryear{Ade \& al}{2014}]{Ade14}
P.A.R. Ade et al, 2014, A \& A 571, A20

\bibitem[\protect\citeauthoryear{Amendola}{2000A}]{Ame00A}
L. Amendola., 2000, Phys. Rev. D 62, 043511

\bibitem[\protect\citeauthoryear{Amendola}{2000B}]{Ame00}
L. Amendola., 2000, MNRAS, 312:521

\bibitem[\protect\citeauthoryear{Amendola}{2004}]{Ame04}
L. Amendola., 2004, Phys.Rev.D69:103524

\bibitem[\protect\citeauthoryear{Amendola \& al}{2014}]{Ame14}
L. Amendola et al, 2014, Phys. Rev. D89, 063538

\bibitem[\protect\citeauthoryear{Aviles}{2014}]{AviCer11}
A. Aviles and J. L. Cervantes-Cota, 2011, Phys. Rev. D84, 083515

\bibitem[\protect\citeauthoryear{Blake}{2012}]{Bla12}
C. Blake, 2012, MNRAS, 425, 405-414

\bibitem[\protect\citeauthoryear{Borges \& al}{2008}]{Bor08}
H. A. Borges et al, 2008, Phys.Rev.D77:043513

\bibitem[\protect\citeauthoryear{C. Contreras \& al}{2013}]{Con13}
C. Contreras et al, 2013, MNRAS, 430, 934-945

\bibitem[\protect\citeauthoryear{A. Costa \& al}{2014}]{Cos14}
A. A. Costa et al, 2014, Phys. Rev. D 89, 103531

\bibitem[\protect\citeauthoryear{Crooks \& al}{2003}]{Cro03}
J. L. Crooks et al, 2003, Astropart.Phys. 20, 361-367

\bibitem[\protect\citeauthoryear{Delubac}{2015}]{Del14}
T. Delubac, 2015, A\&A 574, A59

\bibitem[\protect\citeauthoryear{Devi \& al}{2015}]{Dev15}
N. C. Devi et al, 2015, MNRAS, 448, 37-41

\bibitem[\protect\citeauthoryear{Fay \& al}{2007}]{Fay07}
S. Fay et al, 2007,  Phys.Rev. D76:063504

\bibitem[\protect\citeauthoryear{Garcia-Bellido}{1993}]{Gar93}
J. Garcia-Bellido, 1993, Int.J.Mod.Phys.D2:85-95

\bibitem[\protect\citeauthoryear{Geng}{2015}]{Gen15}
J.-J. Geng, 2015, Eur. Phys. J. C 75, 356

\bibitem[\protect\citeauthoryear{Howlett}{2012}]{How12}
C. Howlett et al, 2012, JCAP, 04, 027

\bibitem[\protect\citeauthoryear{Huterer \& al}{2015}]{Hut15}
D. Huterer et al, 2015, Astroparticle Physics, 63, 23-41

\bibitem[\protect\citeauthoryear{Laureijs \& al}{2011}]{Lau11}
R. Laureijs et al, 2011, ESA/SRE, 12

\bibitem[\protect\citeauthoryear{Lee}{2014}]{Seo14}
S. Lee, 2014, JCAP, 02, 021

\bibitem[\protect\citeauthoryear{Lombriser \& Taylor}{2015}]{Lom15}
L. Lombriser et A. Taylor, 2015, arXiv:1509.08458

\bibitem[\protect\citeauthoryear{Olivares}{2005}]{Oli05}
G. Olivares et al, 2005, Phys.Rev. D71, 063523

\bibitem[\protect\citeauthoryear{Perlmutter \& al}{1999}]{Per99}
S. Perlmutter et al., 1999, Astrophys. J., 517, 565-586

\bibitem[\protect\citeauthoryear{Riess \& al}{1998}]{Rie98}
A. G. Riess et al, 1998, Astron. J., 116, 1009-1038

\bibitem[\protect\citeauthoryear{Setare \& Mohammadipour}{2013}]{Set13}
M. R. Setare et N. Mohammadipour, 2013, JCAP, 01, 015

\bibitem[\protect\citeauthoryear{Suzuki}{2012}]{Suz12}
N. Suzuki, 2012, ApJ 746, 85

\bibitem[\protect\citeauthoryear{Taddei \& Amendola}{2014}]{Tad14}
L. Taddei et L. Amendola, 2014, arXiv:1408.3520

\bibitem[\protect\citeauthoryear{Tocchini \& Amendola}{2002}]{Toc02}
D. Tocchini-Valentini \& L. Amendola, 2002, Phys.Rev. D65, 063508

\bibitem[\protect\citeauthoryear{Tojeiro}{2012}]{Toj12}
R. Tojeiro, 2012, MNRAS, 424, 2339-2344

\bibitem[\protect\citeauthoryear{Wei \& Zhang}{2008}]{Wei08}
H. Wei \& S. N. Zhang, 2008, Phys.Rev.D78:023011

\bibitem[\protect\citeauthoryear{Wetterich}{1995}]{Wet95}
C. Wetterich, 1995, Astron.Astrophys.301:321-328

\bibitem[\protect\citeauthoryear{Yang \& Xu}{2014}]{Yan14}
W. Yang \& L. Xu, 2014, Phys. Rev. D 89, 083517

\end{thebibliography}

\label{lastpage}
\end{document}